\documentclass[twocolumn]{aastex63}
\usepackage[mediumspace,mediumqspace,Grey,squaren]{SIunits}
\usepackage{textcomp}

\graphicspath{{./}{figures/}}

\submitjournal{ApJ}

\shorttitle{Stellar populations in Lindsay-113}
\shortauthors{Li et al.}

\begin{document}

\title{When does the onset of multiple stellar populations in star
 clusters occur-III: No evidence of significant chemical variations in  
 main-sequence stars of NGC 419}

\correspondingauthor{Chengyuan Li}
\email{lichengy5@mail.sysu.edu.cn}

\author{Chengyuan Li} 
\affil{School of Physics and Astronomy, Sun Yat-sen University, Zhuhai 519082, China}

\author{Yue Wang} 
\affiliation{Key Laboratory for Optical Astronomy, National
Astronomical Observatories, Chinese Academy of Sciences, 20A Datun
Road, Beijing 100101, China}

\author{Baitian Tang} 
\affil{School of Physics and Astronomy, Sun Yat-sen University, Zhuhai 519082, China}

\author{Antonino P. Milone}
\affiliation{Dipartimento di Fisica e Astronomia ``Galileo Galilei'', Univ. di Padova, Vicolo dell'Osservatorio 3, Padova, IT-35122}
\affiliation{Istituto Nazionale di Astrofisica - Osservatorio Astronomico di
Padova, Vicolo dell'Osservatorio 5, Padova, IT-35122}

\author{Yujiao Yang}
\affil{Department of Astronomy, School of Physics, Peking University, Yi He Yuan Lu 5, HaiDian District, Beijing 100871, China}

\author{Xin Ji} 
\affiliation{Key Laboratory for Optical Astronomy, National
Astronomical Observatories, Chinese Academy of Sciences, 20A Datun
Road, Beijing 100101, China}

\begin{abstract}
{ Recent studies have revealed that the onset age for the presence of multiple stellar populations 
(MPs) in star clusters seems to correspond to the disappearance of the extended main-sequence turnoff (eMSTO) in young clusters, a pattern associated with stellar rotations. A speculative suggestion 
is that MPs might be caused by the magnetic brake, a stellar evolutionary effect linked to the rotation. 
In this work, we use the young massive cluster NGC 419 as a testbed. We examined if its magnetically baked MS stars would exhibit MPs. Using the deep ultraviolet and visible images observed through the {\sl Hubble Space Telescope}, combined with a specific color index that is sensitive to the nitrogen (N) abundance, we examined if its late G- and K-type MS stars are affected by N variation. Our analysis reports that the morphology of its GK-type MS is most likely an SSP, and only a negligible probability that indicates a N variation up to 0.4 dex is present. We conclude that there is no significant N variation among its GK-type MS stars. The absence of a significant chemical variation among the late-type MS stars indicates that MPs might not be a specific pattern of magnetically braked stars.}
\end{abstract}


\section{Introduction} \label{S1}
Almost all old globular clusters (GCs, older than $\sim$10 Gyr) are composed 
of multiple stellar populations (MPs). One signature of MPs is stars in GCs 
are not chemically homogeneous: they exhibit star-to-star variations in different 
elemental abundances, such as He, C, N, O, Na, Mg, Al, etc
\citep{Piot07a,Yong08a,Carr09a,diCr10a,Lard12a,Wang17a,Panc17a,Milo19a}. 
The number of chemically enriched stars is comparable to, or greater than 
normal stars \citep{Carr09a,Milo12b,Tang17a}. However, such a 
chemical anomaly is rarely detected in field stars. Only a small number of stars with 
similar chemical anomalies are found in the Galactic field, and they are proved 
kinetically related to GCs \citep[e.g.,][]{Tang19a}. MPs seems an exclusive product 
of GCs. 

MPs are detected in extragalactic clusters as well, including clusters in the 
Large Magellanic Cloud, Small Magellanic Cloud \citep[LMC and SMC,][]{Mucc09a,Nied17a}, 
as well as the Fornax dwarf spheroidal galaxy \citep{Lars14a}. For GCs in  
the Milky Way, no significant difference in their MPs was detected between clusters 
associated with different progenitors \citep{Milo20a}. 

About one decade ago, clusters with ages between 1 and 2 Gyr were found to 
harbor an extended main-sequence turnoff (eMSTO) region \citep{Mack08a}. 
The eMSTO was soon proved an ordinary feature for almost all clusters 
younger than $\sim$2 Gyr in the Milky Way and the LMC/SMC \citep{Milo09a,Li17a,Cord18a}. 
The most straightforward explanation to the eMSTO is their host clusters could have  
multiple stellar generations with different ages. However, observations focus on other 
properties of these clusters and their progenitors are inconsistent with this hypothesis \citep{Bast14a,Li14a}. An alternative explanation, stellar fast rotation scenario, 
which suggests that the complex of the turnoff (TO) region is caused by different rotations of 
TO stars \citep{Bast09a,Bran15a,Dant17a}, was proposed and proved a promising explanation 
which accounts for the eMSTO \citep{Mari18a,Sun19a,Sun19b}. 

Although the complexity of MPs is strongly correlated with their host cluster  
mass \citep{Milo17a,Chan19a,Lagi19a}, there seems an onset age for the presence of MPs. 
Almost all GCs older than 10 Gyr, and most clusters with ages between 2 and 10 Gyr, 
exhibit MPs, while all their younger counterparts (younger than 2 Gyr) do not \citep{Mart18a,Li19a,Li19b}. Intriguingly enough, the onset age for the presence of 
MPs also determines the beginning of the disappearance of the eMSTO. The star-to-star 
chemical variations has only been detected among stars that were magnetically 
braked, at which stage their host clusters would not exhibit eMSTOs. A straightforward 
question is whether the MPs and the magnetic brake are related \citep{Bast18a}. 

Although how does the magnetic brake could cause the MPs remains unclear, if it does 
produce the chemical variations among magnetically braked stars, MPs should be 
present in low-mass MS stars, even in young clusters. The mass boundary at which 
a strong magnetic brake works is determined $\sim$1.4 $M_{\odot}$ \citep{Goud18a}, 
indicating that all MS stars later than (or equal to) G-type should be braked by their 
strong surface magnetic field. \cite{Mart17a} have confirmed that there is no significant 
chemical variations among RGB stars of NGC 419. However, NGC 419 exhibits 
an obvious eMSTO region, which implies that both its TO and RGB stars are not 
magnetically braked. Because of this, it is deserve to examine whether the chemical 
inhomogeneous is present among its low-mass MS stars. In this work, we aim to examine 
if one of the most ordinary features of MPs, the star-to-star N variation, is present 
in the late-G to K-type MS stars in the SMC cluster NGC 419. Using the frames that 
are deeply exposed in ultraviolet and visible passbands observed through the 
{\sl Hubble Space Telescope} ({\sl HST}), we have studied their distributions of a specific color 
index which is sensitive to the N abundance. We have compared the observed GK-type 
MS population stars with a simulated simple stellar population (SSP) and a branch of 
synthetic MPs with different degrees of N variation. Our analysis reports that there is 
no significant star-to-star variation in N abundance among the GK-type stars in NGC 419.

The article is organized as follows: in Section 2 we introduce the data reduction. 
We present the method designation and main results in Section 3. The scientific 
discussions about our results and conclusions are present in Section 4.


\section{Data Reduction} \label{S2}
The datasets used in this work were observed through the {\sl HST}'s 
Ultraviolet andVisual Channel of the Wide Field Camera 3 (UVIS/WFC3), 
and the Advanced Camera for Surveys/Wide Field Channel (ACS/WFC). 
The UVIS/WFC3 frames were observed in both the F343N and F438W 
passbands (program ID: GO-15061, PI: N. Bastian). The ACS/WFC frames 
were observed through the F814W passband (program ID: GO-10369, 
PI: J. Gallagher). Our datasets were derived from the photometry of nine frames of 
the UVIS/WFC3 and six frames of ACS/WFC. The detail information of the observational 
frames in each passband is present in Table \ref{T1}.

\begin{table*}
  \begin{center}
\caption{Description of the observations used in this article.}\label{T1}
  \begin{tabular}{l | l l l l l l}\hline
    Rootname      &  Camera	  & Exposure time & Filter & Program ID & PI name \\\hline
    idio01laq	& UVIS/WFC3 & 1450.5 s	 & F343N	& GO-15061 & N. Bastian & \\
    idio01lbq	& UVIS/WFC3 & 1450.5 s	 &  F343N & &  & \\
    idio01ldq	& UVIS/WFC3 & 1450.5 s	 &  F343N & &  & \\
    idio01leq	& UVIS/WFC3 & 1450.5 s	 &  F343N & &  & \\
    idio01l9q	& UVIS/WFC3 & 2924.3 s	 &  F343N & &  & \\
    idio01lgq	& UVIS/WFC3 & 3035.5 s	 &  F343N & &  & \\
    idio01lhq	& UVIS/WFC3 & 3036.0 s	 &  F343N & &  & \\
    idio02daq	& UVIS/WFC3 & 1454.0 s	 &  F438W & &  & \\		
    idio02dbq	& UVIS/WFC3 & 1454.0 s	 &  F438W & &  & \\\hline	
    j96123bxq	& ACS/WFC & 10.0 s	 &  F814W & GO-10396 & J. Gallagher & \\
    j96123bzq	& ACS/WFC & 10.0 s	 &  F814W & &  & \\
    j96123c0q	& ACS/WFC & 474.0 s	 &  F814W & &  & \\
    j96123c2q	& ACS/WFC & 474.0 s	 &  F814W & &  & \\
    j96123c4q	& ACS/WFC & 474.0 s	 &  F814W & &  & \\
    j96123c6q	& ACS/WFC & 474.0 s	 &  F814W & &  & \\\hline
  \end{tabular} 
  \end{center} 
\end{table*} 

The photometry was performed through the package {\sc Dolphot2.0} 
\citep{Dolp11a,Dolp11b,Dolp13a}, a specific photometric package designed 
for {\sl HST} observations. {\sc Dolphot2.0} also contains corresponding 
WFC3 and ACS modules to deal with frames taken from these observational 
channels. We have used the standard photometry routines suggested by the manual. 
The point-spread-function (PSF) photometry was performed to flat frames with poor 
charge transfer efficiency corrected (`{\tt \_flc}'), combined with processes of bad pixel 
masking, splitting frames into different chips, as well as background calculation. 
For each observational channel, our photometry has provided us with a raw stellar 
catalog with parameters including (for each detected object) the positions on the chip, 
$\chi$ (which describes the goodness of the PSF fitting), signal-to-noise (SNR), sharpness, 
roundness, object type, and blocks of photometry in each passband (counts, 
background, count rate, count rate uncertainty, magnitude, magnitude uncertainty etc.). 
We filtered the raw stellar catalog by the following criteria: (1) The object is detected in 
all passbands (otherwise the {\sc Dolphot2.0} will report a magnitude of 99.99); (2) The 
object type is a `bright star' and is not flagged as centrally saturated;  (3) Its sharpness is 
between $-$0.3 and 0.3; (4) Its crowding parameter is less than 0.1 mag. (5) The SNR is 
higher than 5. 

The sharpness describes how sharp a detected source is. A perfectly-fit star should have 
a zero sharpness. A large, positive sharpness means a star is too sharp (perhaps a cosmic 
ray), and a large negative sharpness means the object is too broad (perhaps a blend, cluster, 
or galaxy). In an uncrowded field, good stars should have sharpness values between 
$-$0.3 or 0.3. In a crowded field like a star cluster, blending would affect the sharpness of 
detected stars, making a wider sharpness distribution than field stars. In order 
to improve the reliability of the analysis, we have adopted a strict sharpness range for our 
detections. 

The crowding parameter is in magnitudes, and tells how much brighter the star would have 
been measured had nearby stars not been fit simultaneously. For an isolated star, the value 
is zero. High crowding values are also generally a sign of poorly-measured stars. Crowding 
has a significant effect on our analysis: for a crowded cluster, low-mass stars are usually 
severely affected by crowding, in particular for the cluster's central region. Because of this, 
we have adopted a very strong limitation on crowding for detected stars (not exceed 0.1 mag). 
The average crowding of all GK-type stars analyzed in this work is only 0.02 mag. As a result, a significant fraction ($\sim$50\%) of GK-type stars within the central region (with a size of 
$\sim$300 pixels on the CCD) are removed due to the high crowding.

Thanks to the ultra-deep exposures, in particular for frames observed in the F343N filter 
(the total exposure time is 14,797.8 s). We confirm that in each passband, we can obtain reliable detections down to late-K type stars (F438W$\sim$27 mag) at the distance of the SMC, 
which is sufficient for searching for small N variation ($\delta[N/Fe]\sim0.2$ dex., see Section 3). 

Finally, we combine two stellar catalogs of both the UVIS/WFC3 and 
ACS/WFC channels through cross-matching their spatial coordinates. The 
combined stellar catalog contains totally 23,756 good stars. We have corrected 
the possible effect of differential reddening using the method designed in 
\cite{Milo12a}. We find that for low-mass MS stars, the effect of differential 
reddening is negligible, which will contribute an additional signature of MPs of 
only 1\% or less (see below).

\section{Main Results}
\subsection{Adopted Models}
The color index used for detecting N variation among GK-type stars is identical 
to our previous works \citep[e.g.,][]{Li19b}. Which is: 
\begin{widetext}
\begin{equation}
C_{\rm F343N,F438W,F814W} = {\rm (F343N-F438W)-(F438W-F814W)}
\end{equation}
\end{widetext}
The reason why this color index is sensitive to the N abundance has been illustrated in 
\cite[See their Fig.1]{Li19b}. 

The first thing we did is to find the best fitting isochrone to the observed 
$C_{\rm F343N,F438W,F814W}$ vs. F438W diagram. We have used the MESA 
Isochrone and Stellar Tracks \citep[MIST;][]{Paxt11a,Paxt13a,Paxt15a,Choi16a,Dott16a}. 
The best fitting isochrone is determined by visual inspection, with the best fitting 
parameters of [Fe/H]=$-$0.70$\pm$0.05 dex, $A_{V}$=0.15$\pm$0.01 mag, 
$\log{(t/{\rm yr})}$=9.15$\pm$0.02 ($\sim$1.4 Gyr), $(m-M)_0$=19.0$\pm$0.05 mag, 
respectively. The adopted rotational velocity for the best fitting isochrone is zero, as 
low-mass stars are not fast rotators. The uncertainty associated with each parameter 
is determined by the generated grid when we check the fitting. { By inspecting its 
color-magnitude diagram in optical passbands (F555W and F814W), we find that 
NGC 419 exhibits a very tight RGB, which can constrain the overall metallicity spread. 
We confirm that with a fixed age and distance modulus, the internal spread of [Fe/H] would 
not exceed 0.1 dex. This is well consistent with our fitting uncertainty.} To avoid any 
uncertainty introduced by the fitting, the best fitting isochrone is only used for generating 
N-enriched models, and is not used as the ridge-line of the observed low-mass MS stars. 

We then generate a series of model spectra using the stellar line analysis program 
MOOG \citep[2017 version,][]{Sned73a} and spherical MARCS model atmospheres 
\citep{Gust08a}. Based on the best-fitting isochrone combined with the synthetic 
model spectra, we calculated the corresponding loci with different degrees of N 
enrichment (MPs). { In principle, different CNO abundances will affect stellar evolution 
as well, because they will change the stellar central mean molecular weight as well as 
the atmospheric opacity. We have examined this effect for late-type MS stars of 
different initial masses through MESA. We find that the differences of $\log g$ and 
$\log{T_{\rm eff}}$ between normal and CNO enhanced stars are negligible, 
which are smaller than 0.8\textperthousand$\;$and 1.5\textperthousand, respectively.}

\begin{figure}[!ht]
\includegraphics[width=1\columnwidth]{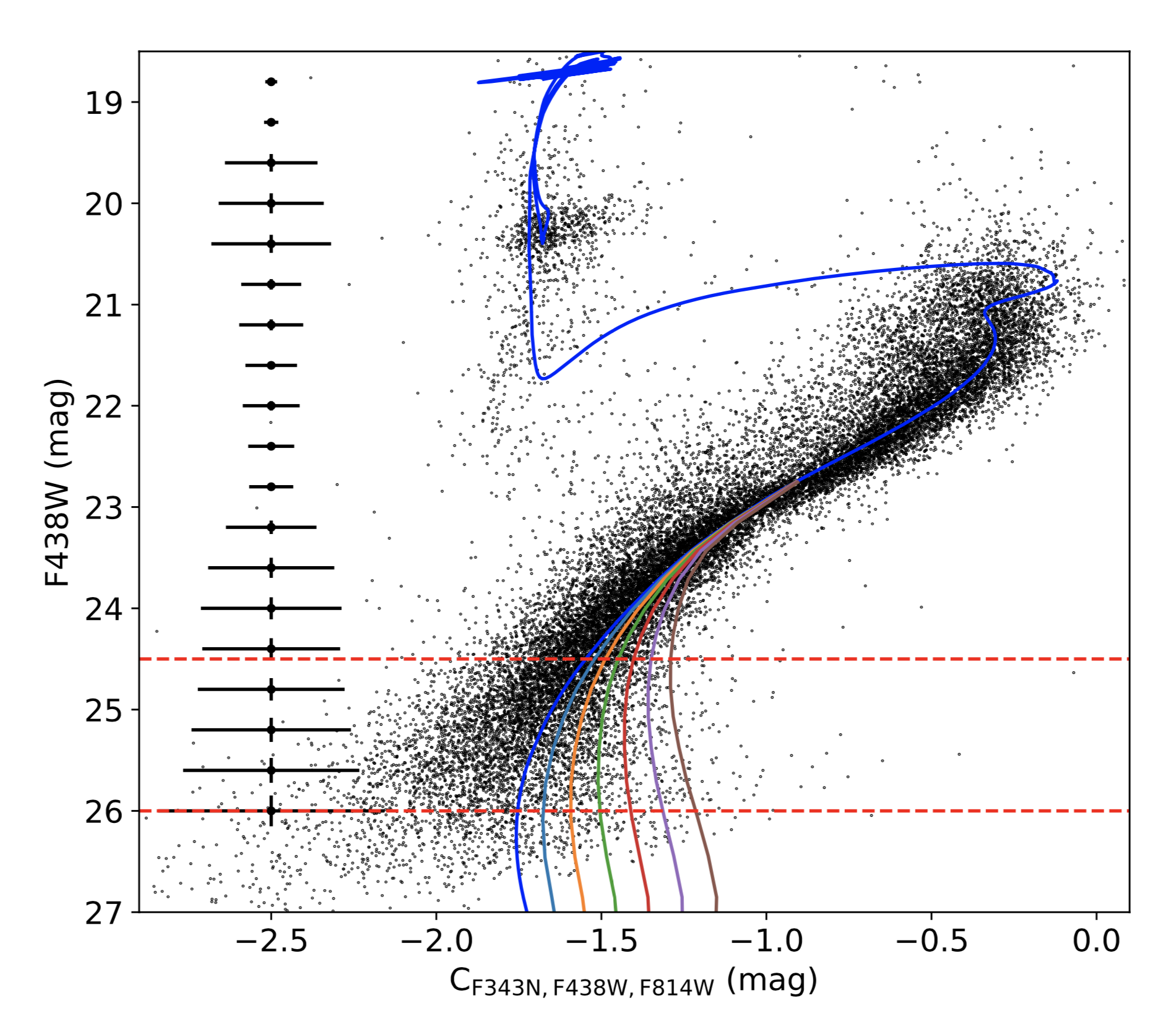}
\caption{The $C_{\rm F343N,F438W,F814W}$ vs. F438W diagram for all stars in the field of 
NGC 419, the best fitting isochrone (blue curve) as well as the MS loci with 
different N-enrichment (color-coded) are present. From left to right, the curves 
are loci with $\Delta$[N/Fe]=0.2, 0.4, 0.6, 0.8, 1.0 and 1.2 dex ($\Delta$[C/Fe]=$\Delta$[O/Fe]=$-$0.1, $-$0.2, $-$0.3, $-$0.4, $-$0.5 and $-$0.6 dex), respectively. Only the MS parts of loci with 
different $\Delta$[N/Fe] is present. MS stars located between the red dashed lines 
are analyzed in this work. {1$\sigma$ level uncertainties for stars of analysis 
are indicated by the error bars on the left.}}
\label{F1}
\end{figure}

All points at the loci have the same global parameters 
($\log g$, $\log{T_{\rm eff}}$, $X,Y,Z$) to the corresponding 
points at the standard isochrone. We calculate the relative deviation 
between loci with $\Delta$[N/Fe]=0.2, 0.4, 0.6, 0.8, 1.0 and 1.2 dex to the 
standard isochrone. An ordinary feature for the N-enriched stars in GCs 
is they are also depleted in C and O, with their total abundance unchanged \citep{Piet09a}. 
To simplify our calculation, we have considered a `toy' model with $\Delta$[C/Fe]+$\Delta$[N/Fe]+$\Delta$[O/Fe]$=$0.0, and $\Delta$[C/Fe]=$\Delta$[O/Fe]=$-$0.5$\Delta$[N/Fe]. 
{ These depletions of C and O abundances were taken into account in our synthesis spectrum.
We confirmed that small variations on $\Delta$[C/Fe] and $\Delta$[O/Fe] would not 
strongly affect our results.} 
For late-type stars, the $C_{\rm F343N,F438W,F814W}$ strongly dependents on [N/Fe] 
because they will exhibit a strong NH absorption band centered at 3370$\AA$\footnote{For 
early-type stars, the NH molecules are destroyed because of their high temperature}. As a 
result, the deviation to the standard isochrone for loci with N-enrichments becomes 
obvious in particular for the bottom of the MS. { For N-enriched stars, we calculated 
their flux ratios compared with normal stars with the same global parameters in each passband. 
We then convert these flux ratios into magnitude differences}. In figure \ref{F1} we present 
the observed $C_{\rm F343N,F438W,F814W}$ vs. F438W diagram for NGC 419, the 
best fitting isochrone (the blue curve) as well as the loci with different $\Delta$[N/Fe]. We 
only present the part of MS for loci with different $\Delta$[N/Fe], because the same property 
for stellar populations of  RGB stars of NGC 419 has been well studied by \cite{Mart17a} 
(no evidence of MPs was detected among its RGB stars). 

At first glance, we find that for MS stars fainter than F438W=24.5 mag, the 
deviations to the standard isochrone for different N-enriched loci become 
obvious. We thus only selected stars below this magnitude. We have also 
removed all stars fainter than F438W=26 mag to increase the average SNR 
of the stellar sample. This adoption finally yields a stellar catalog with 90\% 
stars have their SNR$>$18 (in F343N passband). Based on the best fitting 
isochrone, stars with 24.5 mag$\leq$F438W$\leq$26 mag would 
have their surface effective temperatures range from $\sim$5500 K to 
$\sim$6200 K, corresponding to stellar types from K-type to late G-type. 
The total number of stars in this magnitude range is 3,070.

To mimic a real observation, we have generated a sample of artificial stars (ASs)
located at the standard isochrone and different N-enriched loci, following a 
Kroupa stellar mass function. The total number of stars in each artificial population 
is 327,500, more than 100 times the real observation. For each AS 
population, a 20\% unresolved binary fraction with a flat mass-ratio distribution is  
adopted \citep{Rube10a}. We note that binary fraction would have a very limited 
effect on our results, because for the bottom of the MS, the photometric 
uncertainty is the dominant factor which affects the MS morphology.

For each artificial population, we divided 3,275 sub-samples containing only 100 
stars in each. We added these sub-samples of ASs into the raw images by using 
the same PSF model that applied to real stars. That means we have repeated this  
procedure for 3,275 times for each population. As a result, we obtained seven 
synthetic SSPs with different N-enrichment (from $\Delta$[N/Fe]=0.0 to 1.2 dex). 
Each synthetic SSP have suffered the same effects of real stars, including the 
photometric uncertainty, the contamination of cosmic ray, the blending effect, etc. We 
have also reduced the ASs using the same criteria as for real stars. The resulting 
AS catalogs have the same distribution of crowding, sharpness, and SNR to the real 
observation. Because of the strong limitation of crowding, our artificial stars also have 
a similar spatial distribution like the real observation in each CCD chip. The average 
crowding for ASs is also 0.02 mag in each passband.  

\begin{figure*}[!ht]
\includegraphics[width=2\columnwidth]{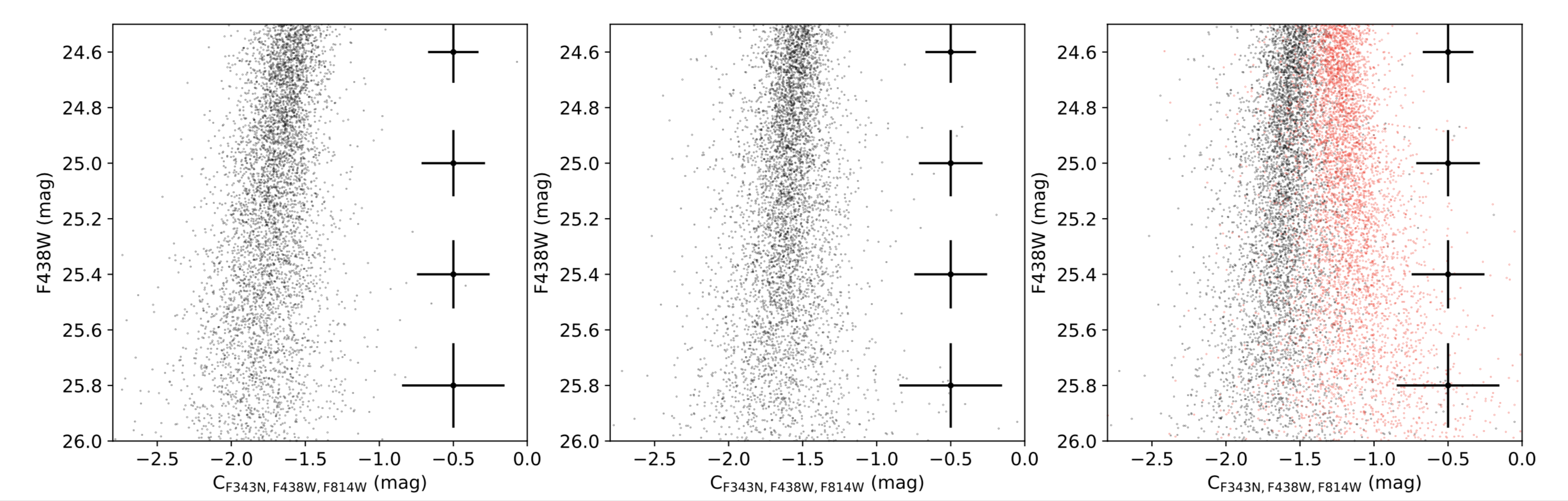}
\caption{The $C_{\rm F343N,F438W,F814W}$ vs. F438W diagrams for the observed GK-type MS stars 
(left), SSP with no N-enrichment (middle) and SSPs with $\Delta$[N/Fe]=0.0 and 1.2 
dex (black and red dots, right). { Error bars are attached on the right of each panel.}}
\label{F2}
\end{figure*}

The photometry of ASs has reported that the average completeness for our stars of 
interest is only 48\%. The incompleteness is mainly contributed by crowding and small 
SNR. The small completeness would not affect our analysis because both the observation 
and the synthetic populations used as comparisons have the same completeness, 
as they are measured through the same PSF photometry and reduced through the 
same criteria. For each population, we randomly selected a represent sample with the same 
number of ASs to the observation based on the observed luminosity function. In figure 
\ref{F2} we present the $C_{\rm F343N,F438W,F814W}$ vs. F438W diagrams for the 
observed GK-type MS stars (left), the synthetic SSP of GK-type MS stars ($\Delta$[N/Fe]=0.0 dex,middle), as well as two synthetic SSPs with $\Delta$[N/Fe]=0.0 and 1.2 dex, respectively 
(black and red dots, right). As shown in figure \ref{F2}, even considering the measurement 
uncertainties, GK-type MS populations with $\Delta$[N/Fe]=0.0 and 1.2 dex are distinct in 
$C_{\rm F343N,F438W,F814W}$. 

\subsection{Statistical Analysis}
\begin{figure*}[!ht]
\includegraphics[width=2\columnwidth]{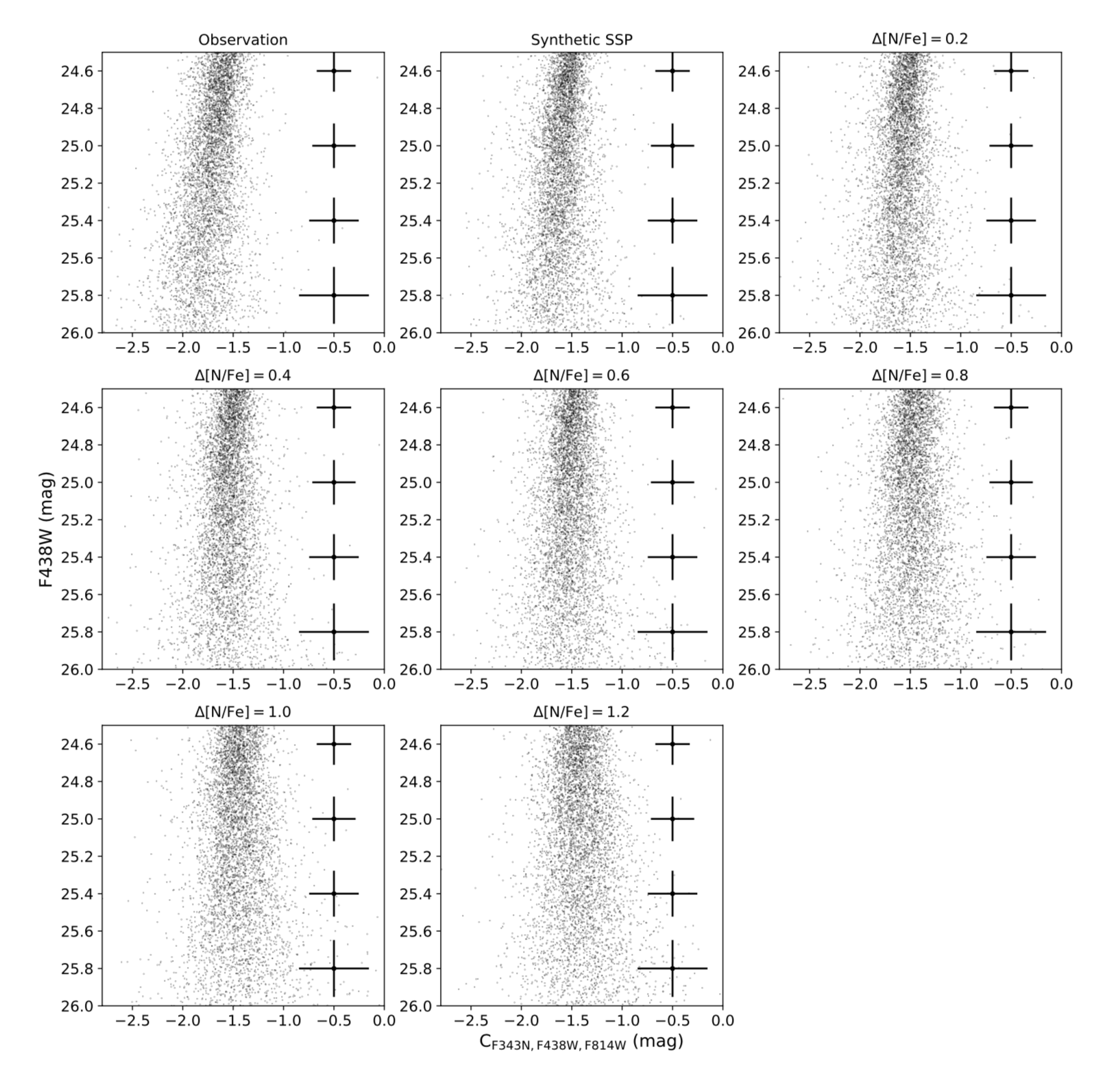}
\caption{The $C_{\rm F343N,F438W,F814W}$ vs. F438W diagrams for the observed GK-type 
MS population, and the corresponding synthetic SSP and MPs (their titles indicate 
the internal variation of N abundance).}
\label{F3}
\end{figure*}

We generate a sample of synthetic MPs through uniformly mixing different SSPs. 
For example, the MP with $\delta$[N/Fe]=0.0 to 0.4 dex would have each 
population ($\Delta$[N/Fe]=0.0,0.2 and 0.4 dex) occupies one third in number. 
In figure \ref{F3} we exhibit the $C_{\rm F343N,F438W,F814W}$ vs. 
F438W diagrams for the observation, the synthetic SSP, and synthetic MPs with 
different N-enrichments. Simply from visual inspection, we can hardly tell the 
difference between the observation and MPs with small N variation 
($\delta{\rm [N/Fe]}\leq0.4$ dex), but a clear difference appears between the observed 
GK-type MS and MPs with $\delta{\rm [N/Fe]}\geq0.6$ dex. 

\begin{figure*}[!ht]
\includegraphics[width=2\columnwidth]{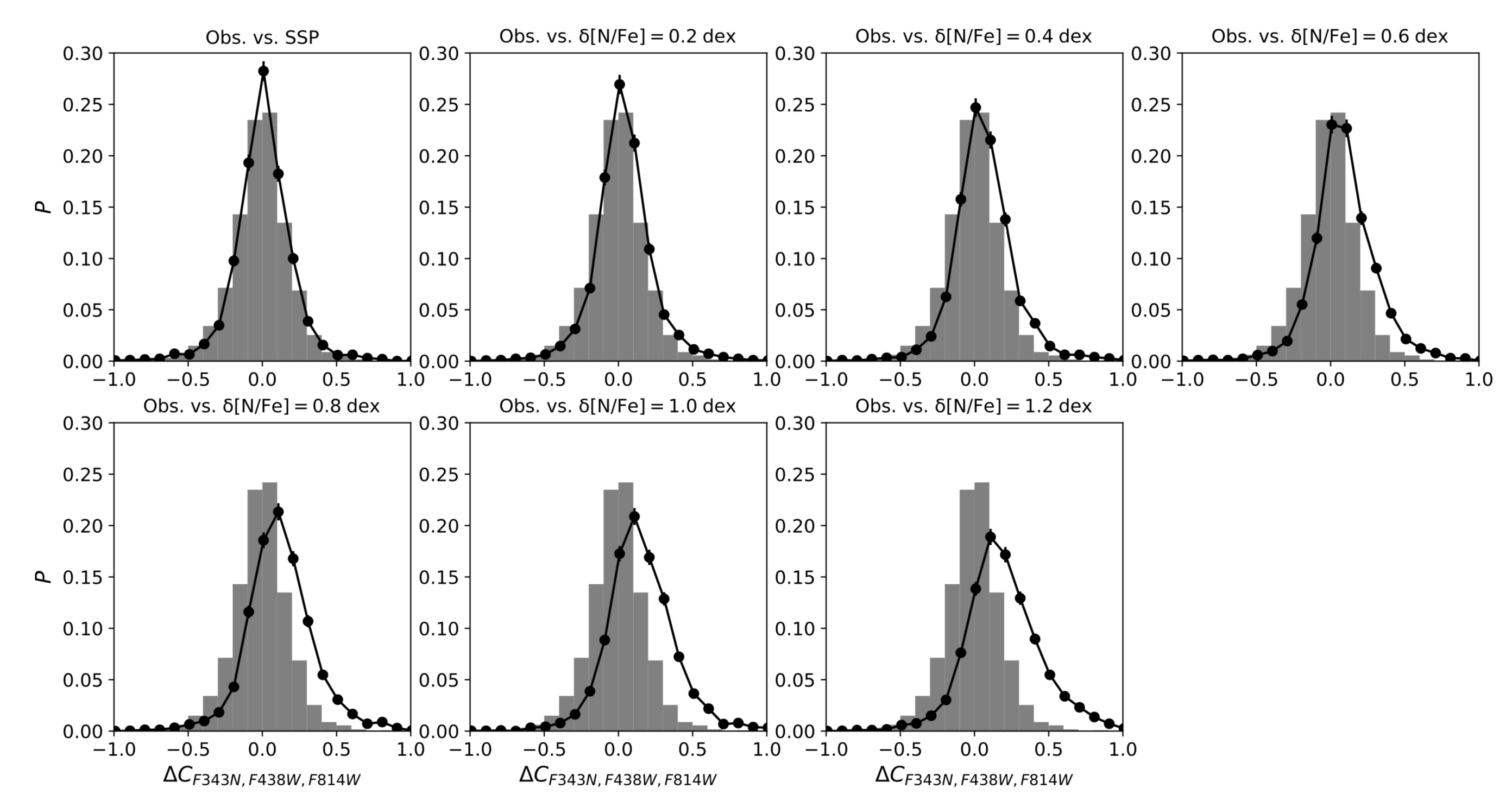}
\caption{Comparisons between the observed $\Delta{C_{\rm F343N,F438W,F814W}}$ 
distribution (grey histograms) and synthetic stellar populations with different N variations (line 
connected black circles)
}
\label{F4}
\end{figure*}

To quantify the differences between the observation and different synthetic  
populations, we have calculated the distribution of $\Delta{C_{\rm F343N,F438W,F814W}}$ 
for both the observed stars and the synthetic ASs. Here $\Delta{C_{\rm F343N,F438W,F814W}}$ 
is the deviation of the detected ${C_{\rm F343N,F438W,F814W}}$ to the MS ridge-line. For 
the observation, the MS ridge-line is determined through connecting the median ${C_{\rm F343N,F438W,F814W}}$ value at different F438W magnitude range with a length of 0.1 
mag. For the synthetic SSP and MPs, their ridge-lines are the best fitting isochrone. 
We then compare the observed $\Delta{C_{\rm F343N,F438W,F814W}}$ distribution with 
different  synthetic stellar populations. Our result is present in figure \ref{F4}. We find that the 
best fitting model to the observation is the SSP. For MPs with $\delta{\rm [N/Fe]}\geq0.4$, a clear displacement to the positive side of the $\Delta{C_{\rm F343N,F438W,F814W}}$ appears. 
This is under expected, because N-enrichment increases the $\Delta{C_{\rm F343N,F438W,F814W}}$ 
of stars due to the strong NH- absorption band covered by the F343N passband. 

Finally, we calculate the standard deviation, $\sigma_{c}$, of the 
$\Delta{C_{\rm F343N,F438W,F814W}}$ distributions for both the observation 
and the synthetic population stars. Considering that the synthetic population 
stars (3,070 stars) were randomly selected from their parent samples (327,500 stars), for 
the synthetic population stars, we have repeated this procedure100 times. We adopt 
the average as the typical value of $\sigma_{c}$. The uncertainty of $\sigma_{c}$ 
is determined by the range of $\sigma_{c}$ of 100 runs. For each model, if $m$ runs have 
produced a $\sigma_{c}$ smaller than or equal to the observation, the probability of that the 
observation can be reproduced by the model is $P$=$m$\%. The $\sigma_{c}$ for the 
observation as well as the synthetic population stars are summarized in Table \ref{T2} 
(second to third columns). 

\begin{table}
  \begin{center}
\caption{The standard deviation of $\Delta{C_{\rm F343N,F438W,F814W}}$ for the observation 
and the synthetic populations. The second and third columns are for `extreme' cases, and 
the forth and fifth columns are for `moderate' cases, see text.}\label{T2}
  \begin{tabular}{l | l l l l l}\hline
    Stellar samples      &  $\sigma_{c}$ (mag)	& $P$ &  $\sigma'_{c}$ (mag) & $P'$  \\\hline
    Observation	&	0.200 & -- & -- & -- \\\hline
    Synthetic SSP & $0.202^{+0.021}_{-0.013}$ & 47\% & $0.202^{+0.021}_{-0.013}$ & 47\%  \\\hline
    Synthetic MPs &  \\
    $\delta$[N/Fe]=0.2 dex & $0.208^{+0.018}_{-0.014}$ & 9\% & $0.208^{+0.031}_{-0.012}$ & 10\%\\
    $\delta$[N/Fe]=0.4 dex & $0.217^{+0.021}_{-0.014}$ & 0\% & $0.213\pm0.018$ & 3\%\\
    $\delta$[N/Fe]=0.6 dex & $0.227^{+0.018}_{-0.015}$ & 0\% & $0.218^{+0.014}_{-0.012}$ & 0\%\\
    $\delta$[N/Fe]=0.8 dex & $0.241^{+0.020}_{-0.012}$ & 0\% & $0.225^{+0.025}_{-0.015}$ & 0\%\\
    $\delta$[N/Fe]=1.0 dex & $0.255^{+0.021}_{-0.017}$ & 0\% & $0.235\pm0.013$ & 0\%\\
    $\delta$[N/Fe]=1.2 dex & $0.276^{+0.026}_{-0.017}$ & 0\% & $0.248^{+0.045}_{-0.017}$ & 0\%\\\hline
  \end{tabular} 
  \end{center} 
\end{table} 

From Table \ref{T2} we can see that the observed $\Delta{C_{\rm F343N,F438W,F814W}}$ 
distribution for GK-type MS stars is most likely the result of an SSP. Among the 100 runs 
of the simulated SSPs, totally 47 runs have reproduced a narrower distribution of 
$\Delta{C_{\rm F343N,F438W,F814W}}$ than the observation ($P$=47\%). 
For the MPs with $\delta$[N/Fe]=0.2 dex, this probability decreases to 9\%. 
For other synthetic MPs with $\delta$[N/Fe]$\geq$0.4 dex, none of them can reproduce 
the observed $\Delta{C_{\rm F343N,F438W,F814W}}$ distribution. 

Some other uncertainties such as the processes of de-reddening, isochrone fitting and 
the adopted binary fraction, would also affect the synthetic population stars. If we do 
not correct the possible differential reddening effect, the observed $\sigma_{c}$ would slightly 
change, from 0.200 to 0.201 mag. If we assume that the late-type MS stars would 
have a lower binary fraction, the synthetic MS population would become slightly narrower. 
But all these effects would not change the fact that the synthetic SSP could have the 
highest probability of reproducing the observation. 

Although we have assumed that the total abundance of the CNO does not change. Some 
other choices of the C and O abundances would affect our results as well, but these effects are 
not significant. This is because for a single star, its F343N magnitude strongly depends on the 
N abundance due to the molecular absorption band of NH ($\sim$3370\AA). The 
F438W magnitude is only weakly affected by the CH absorption band at $\sim$4300\AA\footnote{Because at this wavelength range, the F438W magnitude is 
more sensitive to the continua of the stellar spectral energy 
distribution}. 

One disadvantage of our analysis is we cannot exclude the effect of field contamination. 
Because the field of view of the combined stellar catalog is not large enough for us to 
obtain a referenced field sample. But this should not affect our result, because field stars 
with different ages and metallicities should increase rather than reduce the width of the 
observed MS. The fact that the observed MS is consistent with an SSP should indicate 
that the effect of the field contamination is negligible. 

\begin{figure*}[!ht]
\includegraphics[width=2\columnwidth]{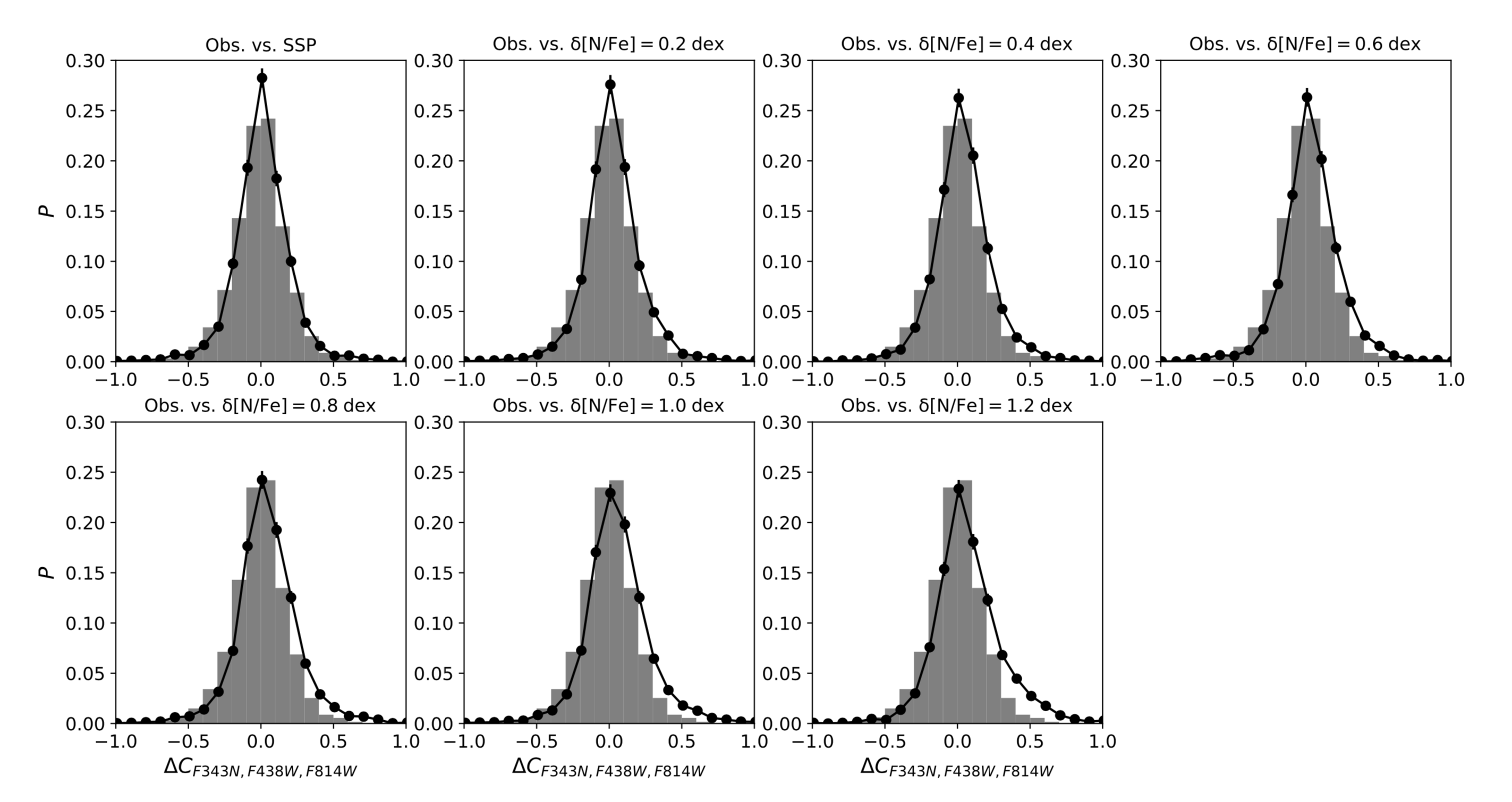}
\caption{The same as figure \ref{F4}, but for the synthetic MPs, the total number fraction 
of N-enriched stars is fixed at 30\%.
}
\label{F5}
\end{figure*}

For all the synthetic MPs in our previous analysis, the number of second population stars 
(stars with $\Delta$[N/Fe]$\geq$0.2 dex) is comparable to, or greater than the primordial 
population stars (stars with no N-enhancement). The fraction of the second population stars in our 
models ranges from 50\% (for MPs with $\delta$[N/Fe]=0.2 dex) to 86\% (6/7, for MPs with 
$\delta$[N/Fe]=1.2 dex), which is somehow extreme compare to real cases of MPs. 
According to \cite{Milo20a}, Magellanic Clouds clusters with MPs could host a smaller 
fraction of second population stars down to $\sim$30\% than Galactic GCs with equivalent 
masses. Because of this, we repeat our previous analysis by adopting some `moderate' 
models of MPs. In this analysis, the number fraction of the primordial population stars is 
fixed as 70\%, and the total number fraction of other population stars with different 
N-enrichment is 30\%. For example, for the model with $\delta$[N/Fe]=1.2 dex, the number 
fractions for the enriched population stars with $\Delta$[N/Fe]=0.2, 0.4, 0.6, 0.8, 1.0 and 1.2 
dex are both 5\% (totally 30\%). Based on this adoption, the same comparisons between the 
$\Delta{C_{\rm F343N,F438W,F814W}}$ distributions of observation and synthetic 
populations (SSP and MPs with different $\delta$[N/Fe]) are present in figure \ref{F5}.

In figure \ref{F5} we find that for all synthetic MPs, their fittings to the real observation are 
better than previous `extreme' cases, but a clear discrepancy at the positive side of the 
$\Delta{C_{\rm F343N,F438W,F814W}}$ still exist for MPs with $\delta$[N/Fe]$\geq$0.4 
dex. Indeed, our statistical analysis reports that the probability that the observation could 
have an internal variation of $\delta$[N/Fe]=0.4 dex is only 3\%. For MPs with $\delta$[N/Fe]$\geq$0.6 dex, none of them can reproduce the observation. For the model of $\delta$[N/Fe]=0.2 dex, 
this probability slightly increases from 9\% to 10\% compared to previous case. Therefore, 
the observation is still likely an SSP rather than a `moderate' case of MPs. For all these `moderate' models of MPs, we summarize their standard deviations of $\Delta{C_{\rm F343N,F438W,F814W}}$ distributions, and the corresponding probabilities of reproducing the observation in Table \ref{T2} (the forth and fifth columns).

In summary, the observed GK-type MS population is most likely an SSP. Their internal chemical 
variation in N abundance, if present, would not exceed $\delta$[N/Fe]$=$0.2 dex. 

\section{Discussion and Conclusion}
\cite{Lagi19a} have compared the properties of MPs between Magellanic Clouds clusters and 
Galactic GCs. They find that Magellanic Clouds GCs with MPs have smaller RGB width than 
Galactic GCs with similar masses, indicating that the typical chemical spread among 
RGB stars in Magellanic Clouds clusters is smaller than Galactic GCs. According 
to \cite{Mari19a}, for Galactic GCs with MPs, the minimum difference in N abundance 
between the primordial and secondary population stars reaches $\sim$0.3 dex (NGC 6121). 
In Table \ref{T2}, our analysis has reported a possibility of 10\% that 30\% second population  
stars with $\Delta$[N/Fe]$=$0.2 dex could exist in low-mass MS stars of NGC 419, this 
probability cannot be ignored. Therefore, the presence of a weaker signature of MPs (a fraction of 
$\leq$30\% stars with $\Delta$[N/Fe]$\leq$0.2 dex) among the GK-type MS stars in NGC 419, 
cannot be excluded by our analysis. In this work, we conclude that a significant chemical 
variation ($\delta$[N/Fe]$\leq$0.4 dex) among low-mass MS stars in NGC 419 does not 
exist. 

In our model MPs, we did not consider the possible effect of helium spread. MS stars with 
enhanced helium abundance would have a higher temperature than normal stars with  
the same masses due to the increased average molecular weight in the central burning region. 
As a result, the helium-enhanced population would exhibit a bluer color compare to the normal 
MS. If assuming a typical helium variation of $\delta{Y}=0.01$ dex \citep[e.g.,][]{Chan19a}, a 
larger color spread of F438W$-$F814W is expected, producing a wider distribution of $\Delta{C_{\rm F343N,F438W,F814W}}$. This makes the fact that low-mass MS stars of NGC 419 are composed 
of MPs more unlikely. { The overall spread of metallicity ([Fe/H]) may affect the observed MS 
as well. This effect, if present, would indicate that NGC 419 has been polluted by Type II supernova. However, the observed tight RGB observed in its optical color-magnitude diagram has excluded 
this possibility.}

Our result favors the conclusion of \cite{Mart17a}, in which they claimed that there is no 
evidence of MPs among the RGB stars in NGC 419. Since NGC 419 is a massive cluster 
with a total mass ($\sim2\times10^5$ $M_{\odot}$) comparable to most GCs, the absence 
of MPs in NGC 419 indicates the importance of cluster age to the onset of MPs. 
However, age should not be the only factor that determines the presence of MPs. Since 
recently both \cite{Li19b} and \cite{Milo20a} have confirmed that a 4 Gyr-old cluster, Lindsay 
113, does not exhibit MPs. 

\cite{Lagi19a} have derived a clear correlation between the clusters mass and their RGB 
width (thus the significance of MPs) for Magellanic Clouds GCs, which is similar to Galactic 
GCs. However, it is unlikely that NGC 419 is not sufficiently massive to harbor MPs, even 
consider the mass loss due to the internal two-body relaxation and 
external tidal effect. Because at least two older counterparts, NGC 1978 and 
NGC 2121, which have comparable masses ($\sim$1--2$\times10^5$ $M_{\odot}$), and 
are only slightly older ($\sim$2 Gyr) than NGC 419, exhibit signature of MPs 
\citep{Mart17a,Li19a}. Does this indicate that clusters age determines 
the onset of MPs while their mass defines the significance of MPs? More investigations 
on clusters of various ages and masses are required to shed light on this question. 

Although speculative, \cite{Bast18a} suggests that MPs may somehow relate to  
the magnetic brake effect. Because all clusters with eMSTO do not exhibit MPs, it 
seems that only for magnetically braked stars could be able to produce star-to-star 
chemical variations. However, our result has excluded this probability since the GK-type 
stars studied in this work should be all magnetically braked, but they do not exhibit 
any signature of MPs. Our result cannot exclude the possibility that stellar chemical 
anomalies are produced among low mass stars and were present only at their later 
stages. However, this is contrary to some ancient GCs, in which 
non-evolved MS stars were found to have abundance anomalies \citep[e.g.,][]{Bril04a}. 
Therefore, we conclude that the apparent coincidence of the lack of eMSTOs and the 
beginning of MPs is likely just a coincidence. But to make a definite conclusion, more 
studies on clusters at the age limit around magnetic brake are required. 

\acknowledgements 
{ This work is based on observations made with the NASA/ESA Hubble Space Telescope, 
obtained from the data archive at the Space Telescope Science Institute. STScI is operated 
by the Association of Universities for Research in Astronomy, Inc. under NASA contract 
NAS 5-26555.}

C. L. and B. T. acknowledges support from the one-hundred-talent project of Sun Yat-sen 
University. C. L. and Y. W. were supported by the National Natural Science Foundation 
of China under grants 11803048. B. T. gratefully acknowledges support from National Natural 
Science Foundation of China under grant No. U1931102. This work has received funding from
 the European Research Council (ERC) under the European Union's Horizon 2020 research 
 innovation programme (Grant Agreement ERC-StG 2016, No 716082 `GALFOR', PI: Milone,
\url{http://progetti.dfa.unipd.it/GALFOR}); from MIUR through the FARE project R164RM93XW (SEMPLICE, PI Milone) and through the PRIN program 2017Z2HSMF (PI Bedin).


\facilities{{\sl Hubble Space Telescope} (UVIS/WFC3 and ACS/WFC)}
\software{\sc dolphot2.0 \citep{Dolp11a,Dolp11b,Dolp13a}, moog \citep{Sned73a}, marcs \citep{Gust08a}}

\end{document}